\documentclass[12pt]{article}

\usepackage[utf8]{inputenc}
\usepackage{cancel}
\usepackage{amssymb}
\usepackage{slashed}
\usepackage{soul}
\usepackage{tabularx}
\usepackage{xcolor}
\usepackage{booktabs}
\usepackage{subcaption} 
\usepackage{colortbl}
\usepackage{authblk}
\DeclareUnicodeCharacter{00A0}{ }

\newcolumntype{C}{>{\centering\arraybackslash}X}
\usepackage[T1]{fontenc}
\usepackage{epsfig}
\usepackage{latexsym}
\usepackage{graphicx}
\usepackage{amsmath}
\usepackage{amsfonts}   
\usepackage{amssymb}    
\usepackage{float}
\usepackage{bm}
\usepackage{url}
\usepackage{hyperref} 
\usepackage[nodisplayskipstretch]{setspace}
\def\order#1{\ensuremath{{\cal O}(#1)}}
\def\lsim{\raise0.3ex\hbox{$\;<$\kern-0.75em\raise-1.1ex\hbox{$\sim\;$}}}
\def\gsim{\raise0.3ex\hbox{$\;>$\kern-0.75em\raise-1.1ex\hbox{$\sim\;$}}}

\def    \beq            {\begin{equation}}
\def    \eeq            {\end{equation}}
\def    \bea           {\begin{eqnarray}}
\def    \eea           {\end{eqnarray}}

\def \mn{\mu\nu{\rm SSM}}

\def\g2{{\rm GeV}^2}

\def\sw2{sin^2 \theta_w}

\def\a^tau{\alpha_{\tau}}

\def\beq{\begin{equation}}
\def\eeq{\end{equation}}
\def\beqa{\begin{eqnarray}}
\def\eeqa{\end{eqnarray}}

\newcommand{\tev}{\,\textrm{TeV}}
\newcommand{\gev}{\,\textrm{GeV}}

\newcommand{\newc}{\newcommand}
\newc\BR{BR}
\newc{\akappa}{A_{\kappa} }
\newc\deltagmtwo{\delta (g-2)_{\mu}} 
\newc\deltaamu{\Delta a_{\mu}}

\def\anti{\overline}

\def\la{\lambda}

\def\ka{\kappa}

\newc{\haa}{BR\(h_1\to a_1 a_1\)}
\newc{\abb}{BR\(a_1\to b\anti{b}\)}
\newc{\hbb}{BR\(h_1\to b\anti{b}\)}
\newc{\abund}{\Omega h^2}
\newc\bsgamma{b\rightarrow s \gamma }
\newc\bxsgamma{\overline{B}\rightarrow X_{s}\gamma}
\newc\brbsgamma{\BR(\overline{B}\rightarrow X_s\gamma)}

\newc{\Fermi}{\textit{Fermi}-}

\allowdisplaybreaks

\usepackage{feynmp}
\graphicspath{{./Figures/Diagrams/}}
\usepackage{enumitem}

\setul{0.5ex}{0.3ex}
\definecolor{Blue}{rgb}{0,0.0,1}
\setulcolor{Blue}

\usepackage{array, makecell}
\usepackage{boldline}
\usepackage[nosort]{cite}

\title{{\bf Searching for a squark LSP of the first two families at the LHC
}}

\author[a]{Paulina~Knees\thanks{pknees@df.uba.ar}}
\author[b,c]{Essodjolo Kpatcha\thanks{kpatcha.essodjolo@uam.es}}
\author[d]{I\~naki Lara\thanks{inaki.lara@fuw.edu.pl}}
\author[a,e]{Daniel~E.~L\'opez-Fogliani\thanks{daniel.lopez@df.uba.ar}}
\author[b,c]{Carlos~Mu\~noz\thanks{c.munoz@uam.es}} 

 \affil[a]{Instituto de Física de Buenos Aires UBA \& CONICET, Departamento de Física, Facultad de Ciencia Exactas y Naturales, Universidad de Buenos Aires, 
 1428 Buenos Aires, Argentina}
 \affil[b]{Departamento de F\'{\i}sica Te\'{o}rica, Universidad Aut\'{o}noma de Madrid (UAM),
Campus~de~Cantoblanco, 28049 Madrid, Spain}
  \affil[c]{Instituto de F\'{\i}sica Te\'{o}rica (IFT) UAM-CSIC, Campus de Cantoblanco, 28049 Madrid, Spain}
  \affil[d] 
  {Faculty of Physics, University of Warsaw, Pasteura 5, 02-093 Warsaw, Poland}
 \affil[e]{
 {Pontificia Universidad Católica Argentina, 
 Av. Alicia Moreau de Justo~1500, 
 1107~Buenos~Aires, Argentina}}

\date{}

\begin{document}
\maketitle
\begin{abstract}
We analyse relevant signals expected at the LHC for a squark of the first two families as the lightest supersymmetric particle (LSP).
The discussion is established in the framework of the $\mu\nu$SSM, where the presence of $R$-parity violating couplings involving right-handed neutrinos solves simultaneously the $\mu$-problem and the accommodation of neutrino masses and mixing angles.
The squarks are pair produced and decay dominantly to a neutrino and a quark. They also have two sub-dominant three body decays to quark, Higgs and neutrino/charged lepton. The decays can be prompt or displaced, depending on the regions of the parameter space of the model.
We focus the analyses on squarks of right up-type, right down-type, and left up-type; since squarks of left down-type cannot be the LSP because of D-term contributions.
We compare the predictions of these scenarios with ATLAS and CMS searches for prompt and long-lived particles.
To analyse the parameter space
we sample the $\mu\nu$SSM for a squark LSP,
paying special attention to reproduce the current experimental data on neutrino and Higgs physics, as well as flavour observables. Because of the contribution of squark-squark final states to the production cross section, the results depend on the squark family. In particular, for a right strange squark LSP,
the lower limit on the mass is $1646$~GeV, corresponding to an upper limit on the decay length of $54.7$~mm, and 
for a right (left) scharm LSP, the limits are 1625 (1357) GeV and 13.4 (1.9) mm. However, the first family of squarks as LSP turns out to be excluded, unless the gluino is heavier than 7 TeV, which produces a limit on the squark mass of 1800 GeV.
\end{abstract}

\newpage 

\tableofcontents 


\section{Introduction}

The `$\mu$~from~$\nu$' Supersymmetric Standard Model
($\mn$)~\cite{LopezFogliani:2005yw,Escudero:2008jg} (for a review, see Ref.~\cite{Lopez-Fogliani:2020gzo}) is a predictive model alternative to the 
Minimal Supersymmetric Standard Model (MSSM)~\cite{Nilles:1983ge,Barbieri:1987xf,Haber:1984rc,Gunion:1984yn, Martin:1997ns} 
and the Next-to-MSSM (NMSSM)~\cite{Maniatis:2009re,Ellwanger:2009dp}.
In the $\mn$, the presence of couplings involving right-handed (RH) neutrinos solves simultaneously the $\mu$ problem~\cite{Kim:1983dt,Bae:2019dgg} and the $\nu$-problem (the generation of neutrino masses), without
the need to introduce additional energy scales beyond the supersymmetry (SUSY)-breaking scale. In contrast to the MSSM, and the NMSSM, $R$-parity
and lepton number are not conserved,
leading to a completely different
phenomenology characterized by distinct prompt or displaced
decays of the lightest supersymmetric particle (LSP),
producing multi-leptons/jets/photons with small/moderate missing transverse energy (MET) from 
neutrinos~\cite{Ghosh:2017yeh,Lara:2018rwv,Lara:2018zvf,Kpatcha:2019gmq,Kpatcha:2019pve,Heinemeyer:2021opc,Kpatcha:2021nap,Knees:2023fel,Kpatcha:2024imb}.
The smallness of neutrino masses is directly related with the low decay width of the LSP. Actually, it is also related to the existence of possible candidates for decaying dark matter in the model.
This is the case of 
the gravitino~\cite{Choi:2009ng,GomezVargas:2011ph,Albert:2014hwa,GomezVargas:2017,Gomez-Vargas:2019mqk}, or the axino~\cite{Gomez-Vargas:2019vci}, with lifetimes greater than the age of the Universe.
Recently, the role of sterile neutrinos as decaying dark matter candidates in the framework of the $\mn$ has also being studied~\cite{Knees:2022wbt}. 
It is also worth mentioning, concerning cosmology, that baryon asymmetry might be realized in the
$\mn$ through electroweak (EW) baryogenesis~\cite{Chung:2010cd}. The EW sector of the $\mn$ can also explain~\cite{Kpatcha:2019pve,Heinemeyer:2021opc} the longstanding
discrepancy between the experimental result for the anomalous magnetic
moment of the muon~\cite{Abi:2021gix,Albahri:2021ixb} and its Standard Model (SM) prediction~\cite{Aoyama:2020ynm}.\footnote{
In this work we will not try to explain it
since we are interested in the analysis of a squark LSP through the decoupling of the rest of the SUSY spectrum.}

Because of $R$-parity violation (RPV) in the $\mn$, basically all SUSY particles are candidates for the LSP, and therefore analyses of the LHC phenomenology associated to each candidate are necessary to test them. Given the current experimental results on SUSY search, this is clearly a crucial task. Although this task was first concentrated on the EW sector of the $\mn$, analysing left sneutrinos, the right smuon and the bino as candidates for the LSP~\cite{Ghosh:2017yeh,Lara:2018rwv,Lara:2018zvf,Kpatcha:2019gmq,Kpatcha:2019pve,Heinemeyer:2021opc}, its colour sector has also been analysed more recently.
In particular, in Ref.~\cite{Kpatcha:2021nap}
the SUSY partners of the top quark as LSP candidates, i.e. the left and right stops, were considered, whereas in  Ref.~\cite{Knees:2023fel} the right sbottom was also considered. 
Recently, in Ref.~\cite{Kpatcha:2024imb} the gluino LSP in the $\mn$ was discussed in detail.

The aim of this work is to finish with the systematic analysis of LSP candidates in the colour sector of the $\mn$, focusing now on squarks of the first two families.
Although we do not concern ourselves with the high-energy origin of this low-energy mass spectrum, it is worth mentioning that it can be obtained simply by assuming non-universal soft terms at high energy.
See also e.g. Refs.\cite{Yamaguchi:2016oqz,Yin:2016shg,Yanagida:2018eho} for a mechanism allowing the first two families of squarks lightest than the stops, even with universal soft sfermion masses.
Thus, we will study the constraints on the parameter space of the model by sampling it to get an
up-type squark or a down-type squark as the LSP in a wide range of masses. We will pay special attention
to reproduce 
{neutrino masses and mixing angles~\cite{Capozzi:2017ipn,deSalas:2017kay,deSalas:2018bym,Esteban:2018azc,deSalas:2020pgw,Esteban:2020cvm}.} 
In addition, we will impose on the resulting parameters agreement with Higgs data as well as with flavour observables.

The paper is organized as follows. In Section~\ref{sec:model}, we will review the $\mn$ and its relevant parameters for our analysis of neutrino, neutral Higgs and squark sectors.
In Section~\ref{sec:squarks},
we will introduce the phenomenology of a squark of the first two families as the LSP, studying its pair production channels at the LHC and its signals. The latter consist of displaced vertices with a quark and a neutrino from a two-body decay mode and a quark, Higgs and neutrino/charged lepton from two three-body sub-dominant decay modes.
In Section~\ref{strategy}, we will discuss the strategy that we will employ to
perform scans searching for points of the parameter space of our scenario compatible with current experimental data on neutrino and Higgs physics, as well as flavour observables such as 
$B$ and $\mu$ decays.
The results of these scans will be presented 
in Section~\ref{sec:results}, and applied to show the 
current reach of the LHC search on the parameter space of a squark of the first two families as the LSP based on ATLAS and CMS 
results~\cite{ATLAS:2022pib,CMS:2019qjk,CMS:2020iwv,CMS:2024trg}.
Finally, 
our conclusions are left for Section~\ref{sec:conclusions}.

\section{The $\mu\nu$SSM}
\label{sec:model}

In the $\mn$~\cite{LopezFogliani:2005yw,Escudero:2008jg,Lopez-Fogliani:2020gzo}, the particle content of the MSSM
is extended by RH neutrino superfields $\hat \nu^c_i$. 
{The simplest superpotential of the model 
is the following~\cite{LopezFogliani:2005yw,Escudero:2008jg,Ghosh:2017yeh}: 
\bea
W &=&
\epsilon_{ab} \left(
Y_{e_{ij}}
\, \hat H_d^a\, \hat L^b_i \, \hat e_j^c +
Y_{d_{ij}} 
\, 
\hat H_d^a\, \hat Q^{b}_{i} \, \hat d_{j}^{c} 
+
Y_{u_{ij}} 
\, 
\hat H_u^b\, \hat Q^{a}
\, \hat u_{j}^{c}
\right)
\nonumber\\
&+& 
\epsilon_{ab} \left(
Y_{{\nu}_{ij}} 
\, \hat H_u^b\, \hat L^a_i \, \hat \nu^c_j
-
\lambda_i \, \hat \nu^c_i\, \hat H_u^b \hat H_d^a
\right)
+
\frac{1}{3}
\kappa_{ijk}
\hat \nu^c_i\hat \nu^c_j\hat \nu^c_k,
\label{superpotential}
\eea
where the summation convention is implied on repeated indices, with $i,j,k=1,2,3$ the usual family indices of the SM 
and $a,b=1,2$ $SU(2)_L$ indices with $\epsilon_{ab}$ the totally antisymmetric tensor, $\epsilon_{12}= 1$. 
}

{Working in the framework of a typical low-energy SUSY, the Lagrangian  containing the soft SUSY-breaking terms related to the superpotential W above,
is given by:
\bea
-\mathcal{L}_{\text{soft}}  =&&
\epsilon_{ab} \left(
T_{e_{ij}} \, H_d^a  \, \widetilde L^b_{iL}  \, \widetilde e_{jR}^* +
T_{d_{ij}} \, H_d^a\,   \widetilde Q^b_{iL} \, \widetilde d_{jR}^{*} 
+
T_{u_{ij}} \,  H_u^b \widetilde Q^a_{iL} \widetilde u_{jR}^*
+ \text{h.c.}
\right)
\nonumber \\
&+&
\epsilon_{ab} \left(
T_{{\nu}_{ij}} \, H_u^b \, \widetilde L^a_{iL} \widetilde \nu_{jR}^*
- 
T_{{\lambda}_{i}} \, \widetilde \nu_{iR}^*
\, H_d^a  H_u^b
+ \frac{1}{3} T_{{\kappa}_{ijk}} \, \widetilde \nu_{iR}^*
\widetilde \nu_{jR}^*
\widetilde \nu_{kR}^*
\
+ \text{h.c.}\right)
\nonumber\\
&+&   
m_{\widetilde{Q}_{ijL}}^2
\widetilde{Q}_{iL}^{a*}
\widetilde{Q}^a_{jL}
{+}
m_{\widetilde{u}_{ijR}}^{2}
\widetilde{u}_{iR}^*
\widetilde u_{jR}
+ 
m_{\widetilde{d}_{ijR}}^2
\widetilde{d}_{iR}^*
\widetilde d_{jR}
+
m_{\widetilde{L}_{ijL}}^2
\widetilde{L}_{iL}^{a*}  
\widetilde{L}^a_{jL}
\nonumber\\
&+&
m_{\widetilde{\nu}_{ijR}}^2
\widetilde{\nu}_{iR}^*
\widetilde\nu_{jR} 
+
m_{\widetilde{e}_{ijR}}^2
\widetilde{e}_{iR}^*
\widetilde e_{jR}
+ 
m_{H_d}^2 {H^a_d}^*
H^a_d + m_{H_u}^2 {H^a_u}^*
H^a_u
\nonumber \\
&+&  \frac{1}{2}\, \left(M_3\, {\widetilde g}\, {\widetilde g}
+
M_2\, {\widetilde{W}}\, {\widetilde{W}}
+M_1\, {\widetilde B}^0 \, {\widetilde B}^0 + \text{h.c.} \right).
\label{2:Vsoft}
\eea

In the early Universe, not only the EW symmetry is broken, but in addition to the neutral components of the Higgs doublet fields $H_d$ and $H_u$ also the left and right sneutrinos $\widetilde\nu_{iL}$ and $\widetilde\nu_{iR}$
acquire a vacuum expectation value (VEV). {With the choice of CP conservation, they develop real VEVs denoted by:  
\begin{eqnarray}
\langle H_{d}^0\rangle = \frac{v_{d}}{\sqrt 2},\quad 
\langle H_{u}^0\rangle = \frac{v_{u}}{\sqrt 2},\quad 
\langle \widetilde \nu_{iR}\rangle = \frac{v_{iR}}{\sqrt 2},\quad 
\langle \widetilde \nu_{iL}\rangle = \frac{v_{iL}}{\sqrt 2}.
\end{eqnarray}
The EW symmetry breaking is induced by the soft SUSY-breaking terms
producing
$v_{iR}\sim {\order{1 \tev}}$ as a consequence of the right sneutrino minimization equations in the scalar potential~\cite{LopezFogliani:2005yw,Escudero:2008jg,Ghosh:2017yeh}.
Since $\widetilde\nu_{iR}$ are gauge-singlet fields,
the $\mu$-problem can be solved in total analogy to the
NMSSM
through the presence in the superpotential (\ref{superpotential}) of the trilinear 
terms $\lambda_{i} \, \hat \nu^c_i\,\hat H_u \hat H_d$.
Then, the value of the effective $\mu$-parameter is given by 
$\mu=
\la_i v_{iR}/\sqrt 2$.
These trilinear terms also relate the origin of the $\mu$-term to the origin of neutrino masses and mixing angles, since neutrino Yukawa couplings 
$Y_{{\nu}_{ij}} \hat H_u\, \hat L_i \, \hat \nu^c_j$
 are present in the superpotential {generating Dirac masses for neutrinos, 
$m_{{\mathcal{D}_{ij}}}\equiv Y_{{\nu}_{ij}} {v_u}/{\sqrt 2}$.}
Remarkably, in the $\mu\nu$SSM it is possible to accommodate neutrino masses
and mixing in agreement with experiments~\cite{Capozzi:2017ipn,deSalas:2017kay,deSalas:2018bym,Esteban:2018azc} via an EW seesaw
mechanism dynamically generated during the EW symmetry breaking~\cite{LopezFogliani:2005yw,Escudero:2008jg,Ghosh:2008yh,Bartl:2009an,Fidalgo:2009dm,Ghosh:2010zi,Liebler:2011tp}. The latter takes place 
through the couplings
$\kappa{_{ijk}} \hat \nu^c_i\hat \nu^c_j\hat \nu^c_k$,
giving rise to effective Majorana masses for RH neutrinos
${\mathcal M}_{ij}
= {2}\kappa_{ijk} {v_{kR}}/{\sqrt 2}$.
Actually, this is possible at tree level even with diagonal Yukawa couplings~\cite{Ghosh:2008yh,Fidalgo:2009dm}.
It is worth noticing here that the neutrino Yukawas discussed above also generate the effective bilinear terms $\mu_i \hat H_u\, \hat L_i $
with $\mu_i=Y_{{\nu}_{ij}} {v_{jR}}/{\sqrt 2}$,
used in the bilinear RPV model (BRPV)~\cite{Barbier:2004ez}.

We conclude, therefore, that the $\mn$ solves not only the
$\mu$-problem, but also the $\nu$-problem, without
the need to introduce energy scales beyond the SUSY-breaking one.

The parameter space of the $\mn$, and in particular the
neutrino, neutral Higgs and squark sectors are 
relevant for our analysis in order to reproduce neutrino and Higgs data, and to obtain in the spectrum a squark of the first or second family as the LSP.
In particular, neutrino and Higgs sectors were discussed in Refs.~\cite{Kpatcha:2019gmq,Kpatcha:2019qsz,Kpatcha:2019pve,Heinemeyer:2021opc}, and we refer the reader to those works for details, although we will summarize the results below. First, we discuss here several simplifications that are convenient to take into account given the large number of parameters of the model.
{Using diagonal mass matrices for the scalar fermions, in order to avoid the
strong upper bounds upon the intergenerational scalar mixing (see e.g. Ref.~\cite{Gabbiani:1996hi}), from the eight minimization conditions with respect to $v_d$, $v_u$,
$v_{iR}$ and $v_{iL}$ to facilitate the computation we prefer to eliminate
the
soft masses $m_{H_{d}}^{2}$, $m_{H_{u}}^{2}$,  
$m_{\widetilde{\nu}_{iR}}^2$ and
$m_{\widetilde{L}_{iL}}^2$
in favour
of the VEVs.
Also, we assume} for simplicity in what follows the flavour-independent couplings and VEVs $\lambda_i = \lambda$,
$\kappa_{ijk}=\kappa \delta_{ij}\delta_{jk}$, and $v_{iR}= v_{R}$. Then, the higgsino mass parameter $\mu$, bilinear couplings $\mu_i$ and Dirac and Majorana masses discussed above are given by:
\bea
\mu=3\la \frac{v_{R}}{\sqrt 2}, \;\;\;\;
\mu_i=Y_{{\nu}_{i}}  \frac{v_{R}}{\sqrt 2}, \;\;\;\;
m_{{\mathcal{D}_i}}= Y_{{\nu}_{i}} 
\frac{v_u}{\sqrt 2}, \;\;\;\;
{\mathcal M}
={2}\kappa \frac{v_{R}}{\sqrt 2},
\label{mu2}    
\eea
where 
we have already used the possibility of having diagonal neutrino Yukawa couplings $Y_{{\nu}_{ij}}=Y_{{\nu}_{i}}\delta_{ij}$ in the $\mn$ in order to reproduce neutrino physics.

\subsection{The neutrino sector}
\label{neutrino}

For light neutrinos, under the above assumptions, one can obtain
the following simplified formula for the effective mass matrix~\cite{Fidalgo:2009dm}:
\begin{eqnarray}
\label{Limit no mixing Higgsinos gauginos}
(m_{\nu})_{ij} 
\approx
\frac{m_{{\mathcal{D}_i}} m_{{\mathcal{D}_j}} }
{3{\mathcal{M}}}
                   \left(1-3 \delta_{ij}\right)
                   -\frac{v_{iL}v_{jL}}
                   {4M}, \;\;\;\;\;\;\;\;
        \frac{1}{M} \equiv \frac{g'^2}{M_1} + \frac{g^2}{M_2},         
\label{neutrinoph2}
  \end{eqnarray}     
where $g'$, $g$ are the EW gauge couplings, and $M_1$, $M_2$ the bino and wino soft {SUSY-breaking masses}, respectively.
This expression arises from the generalized EW seesaw of the $\mn$, where due to RPV the neutral fermions have the flavour composition
$(\nu_{iL},\widetilde B^0,\widetilde W^0,\widetilde H_{d}^0,\widetilde H_{u}^0,\nu_{iR})$.
The first two terms in Eq.~(\ref{neutrinoph2})
are generated through the mixing 
of $\nu_{iL}$ with 
$\nu_{iR}$-Higgsinos, and the third one 
also include the mixing with the gauginos.
These are the so-called $\nu_{R}$-Higgsino seesaw and gaugino seesaw, respectively~\cite{Fidalgo:2009dm}.
One can see from this equation that {once ${\mathcal M}$ is fixed, as will be done in the parameter analysis of Section~\ref{sec:parameter},
the most crucial independent parameters determining {neutrino physics} are}:
\bea
Y_{\nu_i}, \, v_{iL}, \, M_1, \, M_2.
\label{freeparameters}
\eea
Note that this EW scale seesaw implies $Y_{\nu_i}\lsim 10^{-6}$
driving $v_{iL}$ to small values because of the proportional contributions to
$Y_{\nu_i}$ appearing in their minimization equations. {A rough} estimation gives
$v_{iL}\lsim m_{{\mathcal{D}_i}}\lsim 10^{-4}$.

Considering the normal ordering for the neutrino mass spectrum,
and taking advantage of the 
dominance of the gaugino seesaw for some of the three neutrino families, three
representative type of solutions for neutrino physics using diagonal neutrino Yukawas were obtained in 
Ref.~\cite{Kpatcha:2019gmq}.
In our analysis we will use the so-called type 2 solutions, which have the structure
\bea
M>0, \, \text{with}\,  Y_{\nu_3} < Y_{\nu_1} < Y_{\nu_2}, \, \text{and} \, v_{1L}<v_{2L}\sim v_{3L},
\label{neutrinomassess}
\eea
In this case of type 2, it is easy to find solutions with the gaugino seesaw as the dominant one for the third family. Then, $v_{3L}$ determines the corresponding neutrino mass and $Y_{\nu_3}$ can be small.
On the other hand, the normal ordering for neutrinos determines that the first family dominates the lightest mass eigenstate implying that $Y_{\nu_{1}}< Y_{\nu_{2}}$ and $v_{1L} < v_{2L},v_{3L}$, {with both $\nu_{R}$-Higgsino and gaugino seesaws contributing significantly to the masses of the first and second family}. Taking also into account that the composition of the second and third families in the second mass eigenstate is similar, we expect $v_{2L} \sim v_{3L}$. 
In Ref.~\cite{Kpatcha:2019gmq}, a quantitative analysis of the neutrino sector was carried out, with the result that the hierarchy qualitatively discussed above for Yukawas and VEVs works properly. 
See in particular Fig.~4 of Ref.~\cite{Kpatcha:2019gmq}, where
$\delta m^2=m^2_2-m^2_1$ 
versus $Y_{\nu_{i}}$ and $v_{iL}$ is shown for the scans carried out in that work, using
the results for normal ordering from Ref.~\cite{Esteban:2020cvm}. 

We will argue in Section~\ref{sec:results} that the 
other two type of solutions of normal ordering for neutrino physics are not going to modify our results.
The same conclusion is obtained in the case of working with the inverted 
ordering for the neutrino mass spectrum. The structure of the solutions is more involved for this case, because the two heaviest eigenstates are close in mass and the lightest of them has a dominant contribution from the first family. Thus, to choose $Y_{\nu_1}$ as the largest of the neutrino Yukawas helps to satisfy these relations. For the second and third family, a delicate balance between the contributions of $\nu_{R}$-Higgsino and gaugino seesaws is needed in order to obtain the correct mixing angles. In particular, a representative type of solutions for the case of inverted ordering has the structure $M>0$, with
$Y_{\nu_3} \sim Y_{\nu_2} < Y_{\nu_1}$, and $v_{1L}<v_{2L}\sim v_{3L}$.

\subsection{The Higgs sector}
\label{sec:higgs}

The neutral Higgses are mixed with right and left sneutrinos, since
the neutral scalars and pseudoscalars in the $\mn$ have the flavour composition
$(H_{d}^0, H_{u}^0, \widetilde\nu_{iR}, \widetilde\nu_{iL}) $.
Nevertheless, the left sneutrinos are basically decoupled from the other states, since
the off-diagonal terms of the mass matrix are suppressed by the small $Y_{\nu_{ij}}$ and $v_{iL}$.
Unlike the latter states, the other neutral scalars can be substantially mixed.
Neglecting this mixing between 
the doublet-like Higgses and the three right sneutrinos, the expression of the tree-level mass of the { SM-like Higgs} is~\cite{Escudero:2008jg}:
\begin{eqnarray}
m_h^2 \approx 
m^2_Z \left(\cos^2 2\beta + 10.9\
{\lambda}^2 \sin^2 2\beta\right),
\end{eqnarray}
where $\tan\beta= v_u/v_d$, and $m_Z$ denotes the mass of the $Z$~boson.
Effects lowering (raising) this mass appear when the SM-like Higgs mixes with heavier (lighter) right sneutrinos. The one-loop corrections are basically determined by 
the third-generation soft {SUSY-breaking} parameters $m_{\widetilde u_{3R}}$, $m_{\widetilde Q_{3L}}$ and $T_{u_3}$
(where we have assumed for simplicity that for all soft trilinear parameters
$T_{ij}=T_{i}\delta_{ij}$).
These three parameters, together with the coupling $\lambda$ and $\tan\beta$, are the crucial ones for Higgs physics. 
Their values can ensure that the model contains a scalar boson with a mass around $\sim 125 \gev$ and properties similar to the ones of the SM Higgs boson~\cite{Biekotter:2017xmf,Biekotter:2019gtq,Kpatcha:2019qsz,Biekotter:2020ehh}.

In addition, $\ka$, $v_R$ and the trilinear parameter $T_{\kappa}$ in the 
soft Lagrangian~(\ref{2:Vsoft}),
are the key ingredients to determine
the mass scale of the {right sneutrinos}~\cite{Escudero:2008jg,Ghosh:2008yh}.
For example, for $\lambda\lsim 0.01$ they are basically free from any doublet admixture, and using their minimization equations in the scalar potential
the scalar and pseudoscalar masses can be approximated respectively by~\cite{Ghosh:2014ida,Ghosh:2017yeh}:
\bea
m^2_{\widetilde{\nu}^{\mathcal{R}}_{iR}} \approx   \frac{v_R}{\sqrt 2}
\left(T_{\kappa} + \frac{v_R}{\sqrt 2}\ 4\kappa^2 \right), \quad
m^2_{\widetilde{\nu}^{\mathcal{I}}_{iR}}\approx  - \frac{v_R}{\sqrt 2}\ 3T_{\kappa}.
\label{sps-approx2}
\eea

Finally, $\lambda$ and the trilinear parameter $T_{\lambda}$ 
not only contribute to these masses
for larger values of $\lambda$, but
also control the mixing between the singlet and the doublet states and hence, they contribute in determining their mass scales as discussed in detail in Ref.~\cite{Kpatcha:2019qsz}.
We conclude that the relevant parameters
in the {Higgs (-right sneutrino) sector} are:
\bea
\lambda, \, \kappa, \, \tan\beta, \, v_R, \, T_\kappa, \, T_\lambda, \, T_{u_3}, \,  m_{\widetilde u_{3R}},
\, m_{\widetilde Q_{3L}}.
\label{freeparameterss}
\eea
Note
that the most crucial parameters for the neutrino sector~(\ref{freeparameters}) are basically decoupled from these parameters controlling Higgs physics.
This simplifies the analysis of the parameter space of the model,
as will be discussed in 
Section~\ref{sec:parameter}.

\subsection{The squark sector}
\label{squarkmasses}

The mass matrices
of squarks in the $\mn$ include new terms with respect to the matrices of the MSSM~\cite{Escudero:2008jg,Ghosh:2017yeh}. However, these terms are negligible given that they are proportional to the small parameters $Y_{\nu_{ij}}$ and $v_{iL}$.
Thus, in practice, the up-type and down-type squarks eigenstates of the $\mu\nu$SSM coincide basically with those of the MSSM. We discuss both cases below.

\subsubsection{Mass Matrix for Up-Squarks}
\label{sbottom}

In the approximation discussed above,
one obtains the following tree-level mass matrix in the flavour basis ($\widetilde u_{iL}$, $\widetilde u_{iR}$), with the subindex $i=1,2$ denoting the first and second family:
\begin{align}
m_{\widetilde{u}_i}^2= 
  \left( \begin{array}{cc} m^2_{u_i} +
m^2_{\widetilde{Q}_{iL}} + \Delta \widetilde u_L  & m_{u_i} X_{u_i}\\
       m_{u_i} X_{u_i}
 &  m^2_{u_i} + m^2_{\widetilde{{u}}_{iR}} +\Delta \widetilde u_R
         \end{array} \right),
\label{matrixsq2} 
\end{align}
where $m_{u_i}$ is the mass of the $i$-family of up-type quarks,
 $\Delta \widetilde u_{L,R}$ 
denote the D-term contributions,
\bea
\Delta \widetilde u_L = m^2_Z \left( \frac{1}{2} - \frac{2}{3} \sin^2\theta_W \right) \cos 2\beta, \quad
\Delta \widetilde u_R =\frac{2}{3}  m^2_Z \sin^2\theta_W \cos 2\beta,
\label{delta-stop}
\eea
with $\theta_W$ the weak-mixing angle, and
$X_{u_i}$ the left-right mixing term of up-type squarks,
\bea
X_{u_i} = \left(\frac{T_{u_i}}{Y_{u_i}}-\frac{\mu}{\tan\beta}\right).
\eea

As can easily be deduced from Eq.~(\ref{matrixsq2}), the physical masses of up-type squarks are controlled mainly by the value of the soft SUSY-breaking parameters

\bea
m_{\widetilde Q_{iL}}, \, m_{\widetilde u_{iR}}, \, T_{u_i}.
\label{freeparametersstop}
\eea
However, for the first and second family of squarks the trilinear parameters are not relevant compared with the two mass parameters, because $T_{u_i}$  contribute to squark masses through the mixing terms which are suppressed by the masses of up-type quarks. 
Playing with the mass parameters, it is straightforward to obtain the lightest eigenvalue dominated either by the left up-type composition ($\widetilde u_{1,2L}$) or by the 
right up-type composition ($\widetilde u_{1,2R}$). 
Note that in the case of the lightest squark mainly $\widetilde u_{1,2L}$, a small value of the common soft mass $m_{\widetilde Q_{1,2L}}$ makes also $\widetilde d_{1,2L}$ light although slightly heavier than 
$\widetilde u_{1,2L}$ at tree level due to the D-term contribution,
$m_{\widetilde d_{1,2L}}^{2} = m_{\widetilde u_{1,2L}}^2
-m_W^2 \cos 2\beta$ with $\cos 2\beta <0$.

\subsubsection{Mass Matrix for Down-Squarks}

For this type of squarks one has the following tree-level mass matrix in the flavour basis ($\widetilde d_{iL}$, $\widetilde d_{iR}$):
\begin{equation}
    m_{\widetilde{d}_i}^2=\begin{pmatrix}
        m_{d_i}^2+m_{\widetilde{Q}_{iL}}^2+\Delta\widetilde{d}_L & m_{d_i} X_{d_i} \\
        m_{d_i} X_{d_i} & m_{d_i}^2+m_{\widetilde{d}_{iR}}^2+\Delta\widetilde{d}_R
    \end{pmatrix},
\label{eq:matrix-sb}
\end{equation}

where $m_{d_i}$ is the mass of the $i$-family of down-type quarks, $\Delta\widetilde{d}_{L,R}$ denote the D-term contributions,
\bea
\Delta \widetilde d_L = - m^2_Z \left( \frac{1}{2} - \frac{1}{3} \sin^2\theta_W \right) \cos 2\beta, \quad
\Delta \widetilde d_R = - \frac{1}{3}  m^2_Z \sin^2\theta_W \cos 2\beta,
\label{delta-sbot}
\eea

and
$X_d$ the left-right mixing term of down-type squarks,
\begin{equation}
    X_{d_i}= \frac{T_{d_i}}{Y_{d_i}}- \mu\tan\beta.
\label{eq:Xb}
\end{equation}

Thus, the physical masses of down-type squarks are controlled mainly by

\bea
m_{\widetilde Q_{iL}}, \, m_{\widetilde d_{iR}}, \, T_{d_i}.
\label{freeparametersstop2}
\eea
But, similar to the previous case of up squarks, 
for the first and second family of down squarks the trilinear parameters are not relevant compared with the two mass parameters. In this case because $T_{d_i}$ contribute to squark masses through the mixing terms, which are suppressed by the masses of down-type quarks. 
Also, playing with the values of these parameters, it is straightforward to obtain the lightest eigenvalue dominated either by the left down-type composition ($\widetilde d_{1,2L}$) or by the 
right down-type composition ($\widetilde d_{1,2R}$). 
In the case of the lightest squark mainly $\widetilde d_{1,2L}$, a small value of the common soft mass $m_{\widetilde Q_{1,2L}}$ makes $\widetilde u_{1,2L}$ slightly lighter than 
$\widetilde d_{1,2L}$ at tree level due to the D-term contribution discussed above.
Thus, in what follows we will focus on a right down-type squark LSP, for which a low value of $m_{\widetilde d_{1R}}$ or $m_{\widetilde d_{2R}}$ is crucial.

\bigskip

\noindent
In our analysis of 
Section~\ref{sec:results}
we will sample the relevant parameter space of the $\mn$, which contains the independent parameters determining neutrino and
Higgs physics in Eqs.~(\ref{freeparameters}) and (\ref{freeparameterss}).
{Nevertheless, the parameters for neutrino physics $Y_{\nu_i}$, $v_{i}$, $M_1$ and $M_2$ are essentially decoupled from the parameters 
controlling Higgs physics.
Thus, for a suitable choice of the former parameters
reproducing neutrino physics, there is still enough freedom to reproduce in addition Higgs data by playing with 
$\lambda$, $\kappa$, $v_R$, $\tan\beta$, $T_{u_3}$, etc., 
as shown in Refs.~\cite{Kpatcha:2019gmq,Kpatcha:2019pve,Heinemeyer:2021opc}. 
As a consequence, we will not need to scan over most of the latter parameters, relaxing 
our computing task. For this task
{we have employed the 
{\tt Multinest}~\cite{Feroz:2008xx} algorithm as optimizer. {To compute
the spectrum and the observables, we have used {\tt SARAH}~\cite{Staub:2013tta} to generate a 
{\tt SPheno}~\cite{Porod:2003um,Porod:2011nf} version of the model. }
}

\section{Squark LSP Phenomenology}
\label{sec:squarks}

Squarks can be abundantly produced at the LHC (see Refs.~\cite{PhysRevD.31.1581,Beenakker:1996ch}).
Its production at colliders is dominated by QCD processes, since the RPV contributions 
are strongly suppressed in the $\mn$. The pair production of coloured SUSY particles at large hadron colliders has been extensively studied. In particular, the relevant Feynman diagrams, as well as the lowest and higher order processes of squark production in quark and gluon collisions, 
can be found summarized in Ref.~\cite{Beenakker:1996ch}. There, one can see that the most relevant input parameters are squark and gluino masses.

Since we do not expect a significant difference from the values predicted in the MSSM, we make use of Prospino2 \cite{Beenakker:1996ed} to compute NLO cross-sections for the production of squark-antisquark and squark pairs of the first two families. This scenario turns out to be different from the third family of squarks analysed in \cite{Kpatcha:2021nap,Knees:2023fel}. This is because the contribution to the production cross-section of squark pairs is proportional to the corresponding quark density inside the proton, which is negligible for stop and sbottom.
In addition, the production of squark pairs is further supressed by the gluino mass,
since they can only be produced from quark pair initial states.
Thus, only squark-antisquark final states are relevant for the computation of the production cross-section of the third (and also second) family.
However, for the first family of squarks, both types of final states must be taken into account. 

In particular, for our range of interest of squark masses between about 200 GeV and 2000 GeV, and considering a gluino mass of 2700 GeV (see Table~\ref{fixed} below), the production cross-section with proton-proton collisions at $\sqrt s = 13$ TeV
for $\tilde{d}$ is in the range between 28.9 pb and 8.08$\times10^{-5}$ pb, for $\tilde{u}$ between 29.2 pb and 7.42$\times10^{-4}$ pb, and for both $\tilde{s}$ and $\tilde{c}$ between 28.6 pb and 1.01$\times10^{-5}$pb.

\subsection{Decay modes}
\label{subsec:Decaymodes}

For the first and second families of squarks, the dominant decay channel is given by the 
two-body decay to quark and neutrinos, shown in Fig.~\ref{fig:decay-dom}}.
This is qualitatively different from the case of the stop or sbottom LSPs, where the decay width to quark and leptons can also be relevant, as discussed in Refs.~\cite{Kpatcha:2021nap,Knees:2023fel}. The reason is that the latter width is proportional to the square of the quark Yukawa couplings, and therefore for the first two families studied here its contribution turns out to be essentially negligible.

There are also two three-body decays present for the squarks. One to quark, Higgs and neutrinos, 
and another one 
to quark, Higgs and charged leptons, as shown in Figs.~\ref{fig:decay-subdom} left and right, respectively. The latter decay channel is only relevant when a left squark is the LSP, because for right squarks it is suppressed by $Y_{d,u}^2$, and therefore negligible.
Applying the results of Ref.~\cite{Djouadi:2000bx} for the MSSM to our $R$-parity violating model, the dominant terms of the decay widths turn out to be  proportional to $m_{\tilde{q}}^3$, but very much suppressed.
Thus, the branching ratios (BRs) for all these channels are only
10\% (or less) of the total decay width. The remaining 90\% (or more) is given by the channel to quark and neutrinos. This is the channel that we discuss now in more detail.

 \begin{figure}[t!]
     \centering
     \includegraphics[scale=0.55]{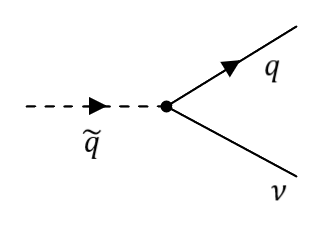}
      \caption{
      Dominant decay channel in the $\mn$ for a squark of the first two families as the LSP.}
     \label{fig:decay-dom}
 \end{figure}

 \begin{figure}[t!]
     \centering
     \includegraphics[scale=0.55]{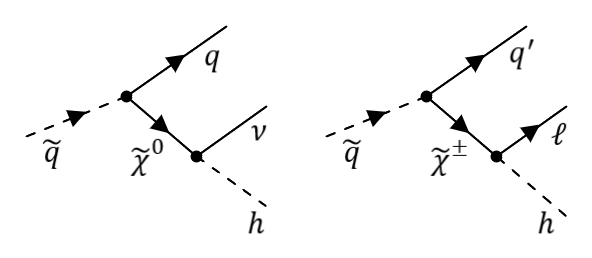}
      \caption{
      Sub-dominant decay channels in the $\mn$ for a squark of the first two families as the LSP. (left) Decay to quark, Higgs and neutrinos; (right) Decay to quark, Higgs and leptons.}
     \label{fig:decay-subdom}
 \end{figure}

The relevant interactions for our analysis of the decay to quark and neutrinos are given in Appendix~\ref{appendix}.
There, one can identify the most important contributions for the decays. 

In particular, for right down-type squarks, the diagram in
Fig.~\ref{fig:decay-dom} corresponds to the second term multiplying the projector $P_{L}$ (and the
first term multiplying the projector $P_{R}$) in 
Eq.~(\ref{neutrinos down}).
It occurs through 
the 
gauge coupling $g'$ of $\widetilde d_{1,2}$
with $d_{1,2}$ and 
neutral bino (Yukawa couplings $Y_{d_{1,2}}$ 
of $\tilde{d}_{1,2}$
with $d_{1,2}$ and neutral higgsinos), via the mixing between bino (higgsinos) and $\nu$.
For right up-type quarks, the diagram corresponds to the first term multiplying the projector $P_L$ in Eq.~(\ref{neutrinos up}). Therefore, it occurs through the gauge coupling $g'$ of $\tilde{u}_{1,2}$ with $u_{1,2}$ and neutral bino, via the mixing between the latter and $\nu$.

For left up-type squarks, the diagram of Fig.~\ref{fig:decay-dom} corresponds to 
the second and third terms multiplying the projector $P_R$
in Eq.~\ref{neutrinos up}. Thus, it occurs similarly to right up-type squarks
but with the addition of the gauge coupling $g$ of $\tilde{u}_{1,2}$ with $u_{1,2}$ and neutral winos, via the mixing between the latter and $\nu$. 

Thus,
in the case of a (pure) right or left squark LSP, the values of the partial decay widths for the different cases can be approximated as: 
\begin{eqnarray}
      \sum_i \Gamma(\widetilde{d}_{1,2R} \to d_{1,2} \nu_i)
      &\sim & \frac{m_{\widetilde{d}_{1,2R}}}{16 \pi} \sum_{i}\left[ \left(\frac{\sqrt{2}}{3}g'U_{i4}^V\right)^2+\left(Y_{d_{1,2}} U_{i6}^V \right)^2\right],
   \label{gamma-bnu}
 \\
    \sum_i \Gamma ({\widetilde u_{1,2R} \to u_{1,2} \nu_i })
    &\sim &
    \frac{
    m_{{\widetilde u}_{1,2R}}}
    {16\pi }
    \sum_i 
 \left( \frac{2\sqrt{2}}{3}g' U^V_{i4} 
 \right)^2,
 \label{eq:up decay}
\\
    \sum_i \Gamma ({\widetilde u_{1,2L} \to u_{1,2} \nu_i }) &\sim& \frac{
   m_{{\widetilde u}_{1,2L}}}  
    {16\pi 
    }\sum_i 
 \left[ \frac{\sqrt{2}}{6}\left(g' U^V_{i4} +  3 g U^V_{i5}
 \right)\right]^2.
 \label{eq:up left decay}
\end{eqnarray}

As discussed in Appendix~\ref{appendix}, $U^V$ is 
the matrix which diagonalizes the mass matrix for the 
neutral fermions, and the above entries 
$U^V_{i4}$, $U^V_{i5}$ and {$U^V_{i6}$}, corresponding to the mixing between neutrinos and bino, winos and neutral higgsino $\widetilde H^0_d$, respectively,
can be approximated as
\begin{eqnarray}
U^V_{i4}\approx\frac{{-}g'}{M_1}\sum_{l}{\frac{v_{lL}}{\sqrt{2}}U^{\text{\tiny{PMNS}}}_{il}},\quad \quad
U^V_{i5}\approx\frac{g}{M_2}\sum_{l}{\frac{v_{lL}}{\sqrt 2}U^{\text{\tiny{PMNS}}}_{il}},\quad \quad
{U^V_{i6}\approx\ \frac{1}{\mu}\sum_l \frac{\mu_l}{\sqrt{2}} U^{\text{\tiny{PMNS}}}_{il}},
\label{--sneutrino-decay-width-2nus2}
\end{eqnarray}
where $U^{\text{\tiny{PMNS}}}_{il}$ are the entries of the PMNS matrix, with
$i$ and $l$ neutrino physical and flavour indices, respectively.
In addition, we use in these formulas $m_{\widetilde{u},\widetilde{d}}\gg m_{u,d}$.
The decay widths given by Eqs.~\ref{gamma-bnu}-\ref{eq:up left decay} are the dominant ones, in good  agreement with the results and figures presented in section~\ref{sec:results}.
Let us remark nevertheless that the results of Section~\ref{sec:results}
have been obtained using the full numerical computation of decay widths,
taking also into account the small contamination between left and right squarks. We have also checked numerically that loop corrections for squark decays are negligible, since the dominant ones are two body decays.

\subsection{LHC searches}
\label{sec:lhc}

The event topologies originated from
a squark LSP of the first two families decaying as described in Subsec.~\ref{subsec:Decaymodes}, produce signals at hadron colliders detectable with diverse LHC searches. As it is shown in Fig.~\ref{fig:decay-dom} and
Fig.~\ref{fig:decay-subdom}, the possible decays include the production of a neutrino and a light quark,  or a neutrino/charged lepton, a light quark and a Higgs boson. Nevertheless, as discussed above, the two body decay dominates. Consequently, the production of a pair of squarks will lead, mostly, to events of the form: $\bar{q}q\bar{\nu}\nu$.
The decay length of the squark LSP ranges from 300~mm scale up to a few mm. Therefore, there are in principle different LHC searches that will have the highest sensitivity for each decay length range. We will classify the signals according to the lifetime scale and apply to each one different searches.

\bigskip
\noindent
{\bf Case i) {Large ionization energy loss}}

\vspace{0.2cm}

\noindent 
Whenever the squark LSP is long-lived enough to measure ionization energy loss and/or time of flight, the searches for heavy charged long-lived particles are able to put strong constraints which are independent of the decay of the squark LSP. 
In particular, the ATLAS search for heavy charged long-lived particles~\cite{ATLAS:2022pib} interprets the analysed data for pair-production of R-hadrons generated from long-lived gluinos. We assume that the hadronization process of a pair of long-lived squarks is similar enough to that of the gluinos to apply the ATLAS results to the squark LSP. 
For each point tested, we compare the predicted cross-section for the production of a pair of squark LSPs with the observed upper limit over the cross-section of the signal of two long-lived gluinos as calculated by ATLAS, for a specific mass and life-time.
If the former is larger than the latter, the point is considered excluded.
\bigskip

\noindent
{\bf Case ii) {Non-prompt jets}}
\vspace{0.2cm}

\noindent 
The timing capabilities of the CMS electromagnetic calorimeter 
 allow discriminating jets arriving at times significantly larger than the travelling times expected for light hadrons, which are moving at velocities close to the speed of light. This time delay can be associated with two effects: First, the larger indirect path formed by the initial trajectory of a long-lived particle plus the subsequent trajectories of the child particles. Secondly, the slower velocity of the long-lived particle due to the high mass compared to light hadrons. Such analysis is performed by the CMS collaboration in the work~\cite{CMS:2019qjk} in the context of long-lived gluinos decaying to gluons and stable gravitinos, excluding gluinos with masses of $\sim2500$ GeV for lifetimes of $\sim1$ m.

The case where the squark LSP decays producing a neutrino and a quark with proper decay lengths above $\sim30$ cm will produce a signal similar to the one analysed in Ref.~\cite{CMS:2019qjk}. For each point analysed in this search we compare the 95\% observed upper limit on cross-section, corresponding to the signal of two delayed jets for a given parent particle mass and $c\tau$, with the prediction of the signal cross-section calculated as $\sigma(pp\to\tilde{q}\tilde{q})\times BR(\tilde{q}\to q \nu)^2$. 
 {Here and in the following, the squark production cross-section is understood as given by the sum of the two processes discussed above producing squark-antisquark and squark pairs.} 
\bigskip

\noindent 
{\bf Case iii) {Displaced vertices}}
\vspace{0.2cm}

\noindent 
For shorter lifetimes, one can confront the points of the model with the limits from events
with displaced vertices, including jets.
The CMS search~\cite{CMS:2020iwv} looks for long-lived gluinos decaying into various final-state topologies containing displaced jets. In particular, it searches for the characteristic signals of long-lived gluino pair production decaying as $\tilde{g}\to g\widetilde{\chi}^0$ or  $\tilde{g}\to    \bar{q} q \widetilde{\chi}^0$. 

These searches originally targets long-lived massive particles with lifetimes in the range 1-10000 mm. Thus, this search can be sensitive to the squark LSP when $c\tau$ is in this range. Some of the signal topologies analysed in the search match exactly the ones originated from the decay shown in Fig.~\ref{fig:decay-dom}. The case where the squark LSP decays producing a neutrino and a quark with proper decay lengths above $\sim1$ mm is contrasted with the model-dependent limits obtained in the ATLAS and CMS searches for displaced vertices. For each point, we compare the 95\% observed upper limit on cross-section, corresponding to the signal of two long-lived gluinos decaying each of them to $g \nu$ for a given parent particle mass and $c\tau$, with the prediction of the signal 
cross-section calculated as $\sigma(pp\to\tilde{q}\tilde{q})\times BR(\tilde{q}\to q \nu)^2$.

The CMS search~\cite{CMS:2024trg} targets similar topologies and proper decay lengths as does~\cite{CMS:2020iwv}, but using a machine learning algorithm to improve the background rejection power. However, the limits obtained for the topologies studied here are equivalent in both searches, thus no improved limit can be obtained for the squark LSP from this analysis.

\section{Strategy for the scanning}
\label{strategy}

In this section, we describe the methodology that we have employed to search for points of our
parameter space that are compatible with the current experimental data on neutrino and Higgs physics, as well as ensuring that a squark of the first two families is the LSP.
In addition, we have demanded the compatibility with some flavour observables,
such as $B$ and $\mu$ decays.
To this end, we have performed scans on the parameter space of the model, with the input parameters optimally chosen.

\subsection{Experimental constraints}

\label{sec:constr}
 
All experimental constraints (except the LHC searches, which are discussed in the previous section) are taken into account as follows:

\begin{itemize}

\item Neutrino observables\\
We have imposed the results for normal ordering from Ref.~\cite{Esteban:2018azc}, selecting points from the scan that lie within $\pm 3 \sigma$ of all neutrino observables. On the viable obtained points we have imposed the cosmological upper
bound on the sum of the masses of the light active neutrinos given
by $\sum m_{\nu_i} < 0.12$ eV~\cite{Aghanim:2018eyx}.

\item Higgs observables\\
The Higgs sector of the $\mn$ is extended with respect to the (N)MSSM.
For constraining the predictions in that sector of the model and address whether a given Higgs scalar of the $\mn$ is in agreement with the signal
observed by ATLAS and CMS, we have interfaced 
{\tt HiggsTools} \cite{Bahl:2022igd} with {\tt Multinest}. For the case of the 
SM-like Higgs boson in the $\mn$, we have used a 
conservative $\pm 3 \gev$ theoretical uncertainty on its mass as obtained with {\tt SPheno}. {\tt HiggsTools} is a unification of HiggsBounds-5~{\cite{Bechtle:2008jh,Bechtle:2011sb,Bechtle:2013wla,Bechtle:2015pma,Bechtle:2020pkv, Bahl:2021yhk}} and
HiggsSignals-2~{\cite{Bechtle:2013xfa,Bechtle:2020uwn}}. 
Our requirement is that the $p$-value reported by {\tt HiggsTools} be larger than 5\%, which is equivalent to impose $\chi^2 < 209$ for the 159 relevant degrees of freedom taken into account in our numerical calculation.

\item $B$~decays\\
$b \to s \gamma$ occurs in the SM at leading order through loop diagrams.
We have constrained the effects of new physics on the rate of this 
process using the average {experimental value of BR$(b \to s \gamma)$} $= (3.55 \pm 0.24) \times 10^{-4}$ provided in Ref.~\cite{Amhis:2012bh}. 
Similarly to the previous process, $B_s \to \mu^+\mu^-$ and  $B_d \to \mu^+\mu^-$ occur radiatively. We have used the combined results of LHCb and CMS~\cite{CMSandLHCbCollaborations:2013pla}, 
$ \text{BR} (B_s \to \mu^+ \mu^-) = (2.9 \pm 0.7) \times 10^{-9}$ and
$ \text{BR} (B_d \to \mu^+ \mu^-) = (3.6 \pm 1.6) \times 10^{-10}$. 
We put $\pm 3\sigma$ cuts from $b \to s \gamma$, $B_s \to \mu^+\mu^-$ and $B_d \to \mu^+\mu^-$, {as obtained with {\tt SPheno}}. {We have also checked that the values obtained are compatible with the $\pm 3 \sigma$ of the recent results from the LHCb collaboration \cite{Santimaria:2021}, and with the combination of results on the rare $B_s^0 \to \mu^+ +\mu^-$ and $B^0 \to \mu^+ +\mu^-$ decays from the ATLAS, CMS, and LHCb \cite{CMS:2020rox}. 

}

\item $\mu \to e \gamma$ and $\mu \to e e e$\\
We have also included in our analysis the constraints {from 
BR$(\mu \to e\gamma) < 4.2\times 10^{-13}$~\cite{TheMEG:2016wtm}}
and BR$(\mu \to eee) < 1.0 \times 10^{-12}$~\cite{Bellgardt:1987du}, {as obtained with {\tt SPheno}}.

\item Chargino mass bound\\
Charginos have been searched at LEP with the result of a lower limit on the lightest chargino mass of 103.5 GeV in RPC MSSM, assuming universal gaugino and sfermion masses at the GUT scale and electron sneutrino mass larger than 300 GeV~\cite{wg1}. This limit is affected if the mass difference between chargino and neutralino is small, and the lower bound turns out to be in this case 92 GeV~\cite{wg2}. LHC limits  can be stronger but for very specific mass relations~\cite{ATLAS:2019lff,CMS:2018xqw,ATLAS:2017qwn,CMS:2018yan}.
Although in our framework there is RPV and therefore these constraints do not apply automatically, we typically choose in our analyses of the $\mn$   a conservative limit of $m_{\widetilde \chi^\pm_1} > 92 \gev$. However, since in this work we are analysing a squark as LSP, the chargino mass is always well above the mentioned bound.

\item Electroweak precision measurements\\
There have been recently several improvements in EW measurements such as $M_W$, $g-2$, $S, T, U$, etc. (see e.g. Refs.~\cite{Heinemeyer:2006px,Chakraborti:2022vds,Cho:2011rk}). Thus, the confrontation of the theory predictions and experimental results might be timely for SUSY models. However, in our framework, electroweak precision measurements are not given significant contributions. This is because the SUSY mass spectrum turns out to be above 1.4 TeV, where the latter value is the lower bound that we obtain in Section~\ref{sec:results} for the mass of a squark LSP.
\end{itemize}

\noindent
Finally, it is worth mentioning that it would not have had a significant impact on the results to have used a criterion of $\pm 2 \sigma$ instead of the one of $\pm 3 \sigma$ used.

\subsection{Parameter analysis}
\label{sec:parameter}
We performed scans using the minimal possible set of parameters necessary to identify points with a squark of the first two families as the LSP. The entire mass spectrum was obtained using the full one-loop numerical computation implemented in \texttt{SPheno}. 
In what follows, we employed the following strategy.
We chose moderate/large values of 
the trilinear coupling $\lambda$ between right sneutrinos and Higgses,
$\lambda = 0.2, 0.4$, thus we are in a similar situation as in the NMSSM (see Ref.~\cite{Domingo:2019jsc} and references therein). Also, small/moderate values of $\tan\beta$, $|T_{u_{3}}|$ and soft stop masses are necessary to obtain through loop effects the correct SM-like Higgs mass~{\cite{Biekotter:2017xmf,Biekotter:2019gtq,Kpatcha:2019qsz,Biekotter:2020ehh}}. For each value of $\lambda$ considered, we used an appropriate range of values of $\tan\beta$ in order to obtain the correct SM Higgs mass, as shown in Table \ref{kap06}.
It is sufficient for our analysis to fix 
$m_{{\widetilde{Q}}_{3L}}$ and $m_{{\widetilde{u}}_{3R}}$ to a reasonable value of
2000 GeV, as shown in
Table~\ref{fixed}, and finally for all the cases we scanned over the low-energy values of $T_{u_3}$ in the range:
\bea
 -T_{u_3} = 900-4500\ \text{GeV}.
 \label{tu}
\eea

In Table~\ref{fixed}, we also show the low-energy values of other input parameters.
Reproducing Higgs data requires suitable additional parameters such as 
$\kappa$, 
$v_R$, $T_\kappa$, $T_\lambda$
(see Eq.~(\ref{freeparameterss})). Thus, we fixed to appropriate value 
$T_\lambda$, which is relevant for obtaining the correct values of the off-diagonal terms of the mass matrix mixing the right sneutrinos with Higgses, and also
$\kappa$, $T_\kappa$, $v_R$ which basically control the right sneutrino sector.
To ensure that chargino is heavier than the squark LSP, the lower value of $\lambda$ forces us to choose a large value for $v_R$ in order to obtain a large enough value of $\mu$ (see Eq.~(\ref{mu2})). As discussed in more detail in the next section, where the parameter analysis is carried out,
smaller values of $\lambda$ imply a chargino LSP.
The parameters $\ka$ and $T_{\kappa}$ are crucial to determine the mass scale of the right sneutrinos.
We choose the value of $-T_{\kappa}$ to have heavy pseudoscalar right sneutrinos,
and therefore 
the value of $\kappa$ has to be large enough in order to avoid 
too light (even tachyonic) scalar right sneutrinos. 
Working with the values of $\lambda$ of Table~\ref{kap06} 
we can keep perturbativity up to an intermediate scale of new physics around $10^{11}$~GeV, as discussed in detail in Ref.~\cite{Kpatcha:2019qsz}.

\begin{table}[t!]
\begin{center}
\begin{tabular}{|c|c|c|}
\hline
 $\lambda$   &0.2 & 0.4 \\
 \hline
$\tan\beta$ & (1.0, 4.0) & (6.5, 8.5)  \\
 \hline
\end{tabular}
\end{center}
\caption{
Low-energy values of the input parameter $\lambda$ determining the two scans carried out, together with an appropriate range of values for $\tan\beta$.
For all the cases, the input parameter $T_{u_3}$ 
is varied in the range shown in Eq.~(\ref{tu}),
and the soft squark LSP mass as discussed in
Eq.~(\ref{sbottommass}).
}
\label{kap06}
\end{table}

\begin{table}[t!]
\begin{center}
\begin{tabular}{|c|}
\hline
$\kappa$ = 0.5\\
$-T_{\kappa}$ = 1000\\
$T_{\lambda}$ = 1500\\
$v_R$ = 3600 \\
$m_{{\widetilde{e}}_{1,2,3R}}=
m_{{\widetilde{Q}}_{3L}} = m_{{\widetilde{u}}_{3R}} = m_{{\widetilde{d}}_{3R}}$ = 2000 \\
$T_{d_{1,2}} = T_{e_{1,2}} = T_{u_{1,2}}$ = 0  \\
$T_{d_3}$ = 100,\ $T_{e_3}$ = 40\\
$-T_{\nu_{1,2,3}}$ = 0.01 \\
$M_1$ = 2400,\ $M_2$ = 2000,\ $M_3$ = 2700 \\
\hline 
\end{tabular}
\end{center}
\caption{Low-energy values of the input parameters that are fixed in the two scans of Table~\ref{kap06}, with the VEV $v_R$ 
and the soft SUSY-breaking parameters given in GeV.
}
\label{fixed}
\end{table}

\begin{table}[t!]
    \centering
    \begin{tabular}{|c|}
    \hline
        $v_{1L} \in (6.3\times10^{-5}, 3.1 \times 10^{-4}
        )$ \\
        $v_{2L} \in (1.2 \times 10^{-4}, 7.9\times 10^{-4})$ \\
        $v_{3L} \in (2.5 \times 10^{-4},  1.0 \times 10^{-3})$ \\
        $Y_{\nu_{1}} \in (3.1 \times 10^{-7},  1.0 \times 10^{-6})$ \\
        $Y_{\nu_{2}} \in (1.2 \times 10^{-6},  6.3 \times 10^{-6})$ \\
        $Y_{\nu_{3}} \in (1.5 \times 10^{-9}, 6.3\times10^{-8})$ \\
        \hline
    \end{tabular}
    \caption{Range of low-energy values of the input parameters related to neutrino physics that are varied in the two scans of Table~\ref{kap06}, with the VEVs $v_{iL}$ given in GeV.}
    \label{tab:neutrino}
\end{table}

The values of other parameters shown in Table~\ref{fixed} concern slepton, squark and  gluino masses, as well as quark and lepton trilinear parameters, which are not specially relevant for our analysis. 
The values chosen for the latter are reasonable within the supergravity framework, where
the trilinear parameters are proportional to the corresponding Yukawa couplings. 
Concerning neutrino physics, as discussed in Section~\ref{neutrino} the most crucial parameters~(\ref{freeparameters}) are basically decoupled from those controlling Higgs physics~(\ref{freeparameterss}).
Thus, for the concrete values of $\lambda$, $\kappa$, $\tan\beta$, $v_R$, etc.,
chosen to reproduce Higgs data, there is still enough freedom to reproduce in addition neutrino data by playing with concrete values
of $M_1$, $M_2$, $Y_{\nu_i}$ and $v_{iL}$, as shown in the last row of Table~\ref{fixed} and
in Table~\ref{tab:neutrino}.

Finally, the soft masses of the squarks, $m_{\widetilde d_{1,2R}}$, $m_{\widetilde u_{1,2R}}$ and $m_{\widetilde Q_{1,2L}}$,
are obviously crucial parameters in our analysis, since they control the physical squark masses of the first two families, as discussed in Sec.~\ref{squarkmasses}.
Thus, e.g., for obtaining a right sdown LSP we scanned its soft mass in the low-energy range
\bea
 m_{\widetilde d_{1R}} = 200-2000\ \text{GeV},
 \label{sbottommass}
\eea
fixing 
$m_{\widetilde d_{2R}}=m_{{\widetilde{u}}_{1,2R}} = m_{{\widetilde{Q}}_{1,2L}}$ = 2000 GeV. For the other possible LSPs, $\widetilde d_{2R}$, $\widetilde u_{1,2R}$, 
$\widetilde u_{1,2L}$, the analyses carried out were similar.

Summarizing, for each $\widetilde q$ LSP we performed two scans over the 9 parameters $m_{\tilde q}$, $T_{u_3}$, $v_{iL}$, $Y_{\nu_i}$ and $\tan\beta$ corresponding to the pair of values of $\lambda$ shown in Table~\ref{kap06}.

\section{Results}
\label{sec:results}

Following the methods described in the previous sections, in order
to find regions consistent with experimental observations we performed scans of the parameter space, and our results are presented here.
To carry this analysis out, we selected first points from the scans that lie within $\pm 3\sigma$ of all neutrino physics observables \cite{Esteban:2018azc}.
Second, we put $\pm 3\sigma$ cuts from $b \to s \gamma$, $B_s \to \mu^+\mu^-$ and $B_d \to \mu^+\mu^-$ 
and require the points to satisfy also the upper limits of $\mu \to e \gamma$ and $\mu \to eee$. 
In the third step, we imposed that Higgs physics is realized.
In particular, we require that the p-value reported by {\tt HiggsTools} be larger
than 5\%.
Also, since we are interested in a squark of the first two families as the LSP, of the allowed points we selected those satisfying this condition.

We show in Fig.~\ref{fig:strange-ctau} the proper decay length for a right strange squark LSP 
for the points of the parameter space studied with $\lambda=0.4$, fulfilling the above experimental constraints. 
As expected, the decay length 
increases with decreasing physical squark masses. As we can see,
all light red points with decay lengths $\gsim 54.7$ mm and squark masses $\lsim 1646$ GeV are excluded. This is due to searches for displaced vertices~\cite{CMS:2020iwv} corresponding to Case (iii).
These searches are also the only ones constraining the right sdown and up-type squarks analysed below.

For the right sdown case, $c\tau$ is basically the same as can be deduced from Eq.~(\ref{gamma-bnu}). The only difference between sdown and strange squark decay widths is due to the contribution proportional to the Yukawa coupling, which is negligible in comparison with the gauge one. However, 
for the second family of squarks the production cross-section is dominated by squark-antisquark final states, as discussed in Sec.~\ref{sec:squarks}, whereas
for the first family, considering the gluino mass as in Table~\ref{fixed}, it is dominated by squark-squark final states. As a consequence, for the sdown case all points turn out to be excluded. One possibility to rescue points, is to increase the gluino mass in order to suppress the squark-squark contribution to the production cross-section. 
In that case, one expects to obtain a similar bound to the one of the strange squark.
We found that for a gluino mass of about 7 TeV, the contribution of both final states to the production cross-section is comparable, recovering points allowed by current constraints and obtaining a lower limit for the mass of the right sdown of 1807 GeV. Increasing the gluino mass up to about 10 TeV makes negligible the squark-squark contribution, and the sdown mass limit becomes similar to the one of the second family of about 1650 GeV.

\begin{figure}[t!]
    \centering
    \includegraphics[width=0.6\textwidth]{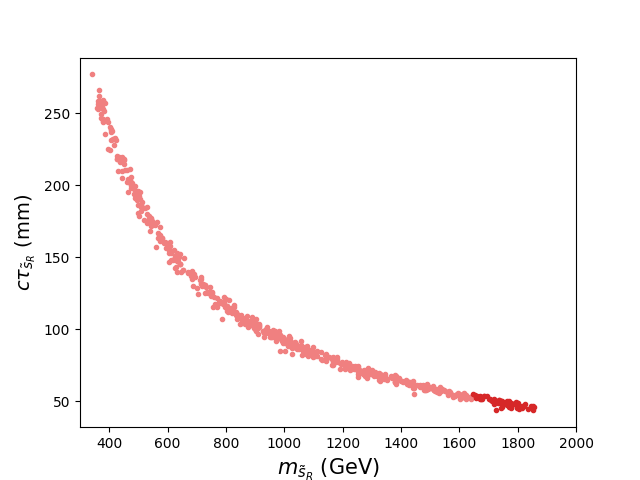}
    \caption{Proper decay length $c\tau$ for a right strange squark LSP versus the right strange physical squark mass. Dark (light) points are allowed (excluded) by LHC  constraints.}
    \label{fig:strange-ctau}
\end{figure}

\begin{figure}[t!]
    \centering
    \includegraphics[width=1.0\textwidth]{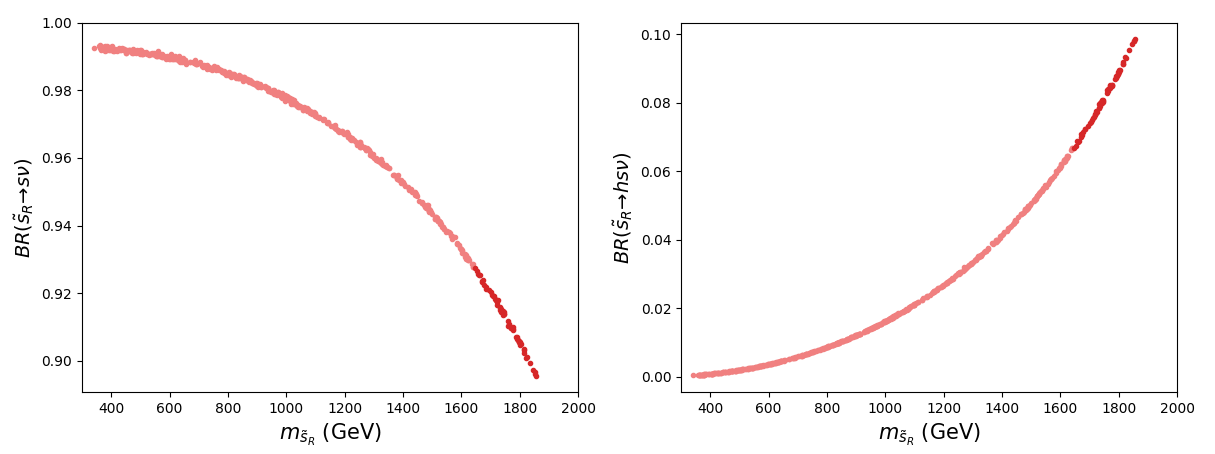}
    \caption{Branching ratios of a right strange squark LSP decaying to $s\nu$ (left plot) and $s h\nu$ (right plot).
    Dark (light) points are allowed (excluded) by LHC constraints. }
    \label{fig:braching-ratios-down}
\end{figure}

The results for the decay length in the case $\lambda = 0.2$ are very similar to the above cases with $\lambda = 0.4$, as expected from Eq.~(\ref{gamma-bnu}).
Besides, given that smaller values of $\lambda$ imply also smaller values of the $\mu$ term for fixed $v_R$ (see Eq.~(\ref{mu2})), the lightest chargino can be lighter than the squarks.
In particular, for physical squark masses larger than about $1500$~GeV we found that the chargino is the LSP. But squark masses smaller than those values are also excluded because $c\tau$ is very close to the one of the case with $\lambda = 0.4$, where the bound for the mass of the right strange is 1646 GeV, as discussed above. Thus, the low $\lambda$ case becomes totally excluded. Although it is possible to increase the $\mu$ term by increasing $v_R$, avoiding the problem of a chargino LSP, again the fact that $c\tau$ is close to the one of the case with $\lambda = 0.4$ makes the mass bound unaffected.

\begin{figure}[t!]
    \centering
    \includegraphics[width=0.6\textwidth]{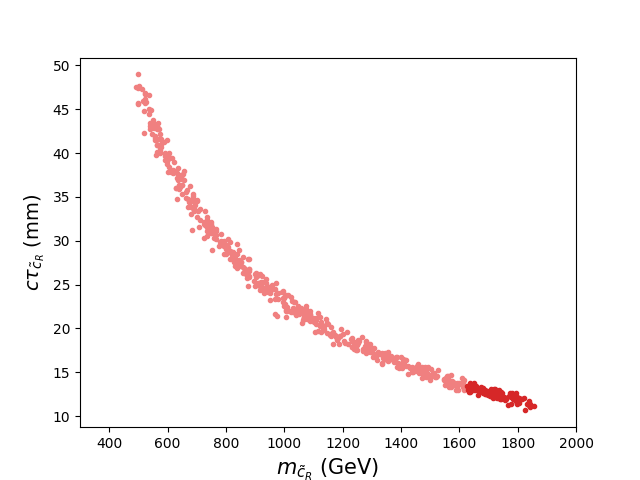}
    \caption{
    The same as in Fig.~\ref{fig:strange-ctau} but for a right scharm LSP.}
    \label{fig:scharm-ctau}
\end{figure}

\begin{figure}[t!]
    \centering
    \includegraphics[width=1.0\textwidth]{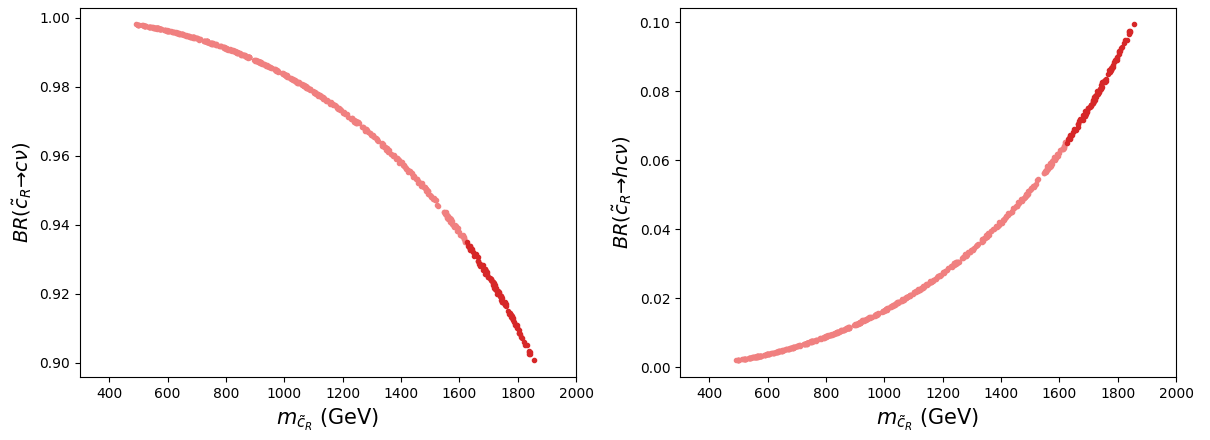}
    \caption{The same as in Fig.~\ref{fig:braching-ratios-down} but for a right scharm LSP. 
    }     
    \label{fig:braching-ratios-scharm}
\end{figure}

In Fig.~\ref{fig:braching-ratios-down}, the BRs of the main decay modes are shown for $\lambda= 0.4$.
As explained in Section~\ref{subsec:Decaymodes}, the sub-dominant channel is the one to $\tilde{q} \to q h \nu $ with a BR $\lesssim 10^{-1}$, implying a sub-dominant contribution to the decay length.

\begin{figure}[t!]
    \centering
    \includegraphics[width=0.6\textwidth]{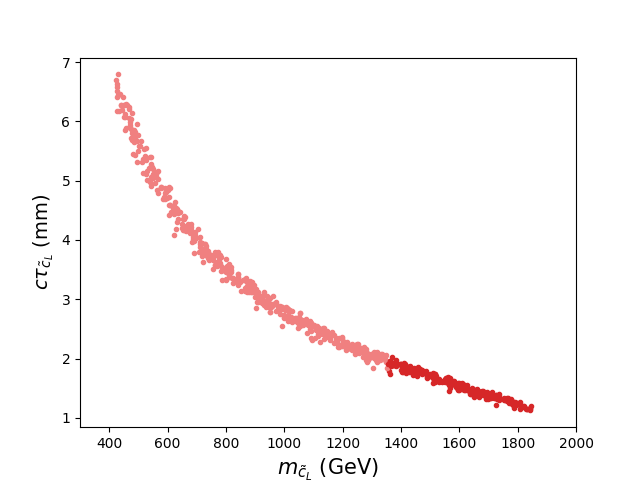}
    \caption{    The same as in Fig.~\ref{fig:strange-ctau} but for a left scharm LSP.
 }
    \label{fig:scharm-left-ctau}
\end{figure}

\begin{figure}[t!]
    \centering
    \includegraphics[width=1.0\textwidth]{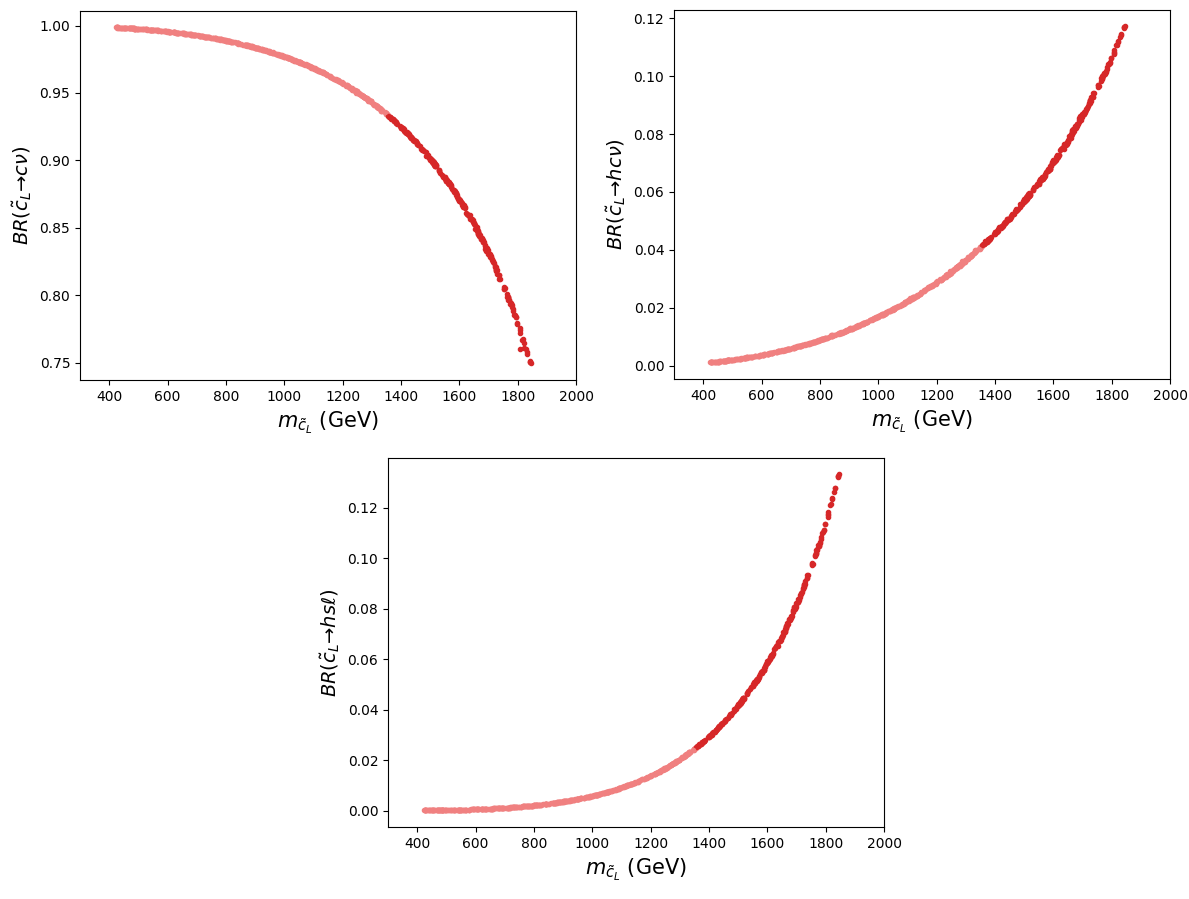}
    \caption{The same as in Fig.~\ref{fig:braching-ratios-down} but for a left scharm LSP including the decay channel to $s h\ell$.}
    \label{fig:braching-ratios-scharm-left}
\end{figure}

The above results can be compared with those of the case of the sbottom LSP analysed in Ref.~\cite{Knees:2023fel}, where the decay width to quark and leptons becomes relevant and is very dependent on $\lambda$. In that case, 
for displaced (prompt) decays, lower limits on the mass of the right sbottom LSP of $1041$~GeV ($1070$~GeV) were obtained. In addition,
the largest possible value found for the decay length is $3.5$~mm.

Regarding the up-type squarks, for a ${\widetilde c_{R}}$ LSP,
similar to the case of the ${\widetilde s_{R}}$ LSP above, the production cross-section is dominated by the squark-antisquark final state.
Thus, we obtained a lower bound on the physical mass of 1625~GeV, which corresponds to an upper bound in the decay length of 13.4~mm. Also, as for the ${\widetilde s_{R}}$, the results are basically independent on $\lambda$, and the same is true for the up-type squarks analysed below. The decay length and BRs obtained are shown in Figs.~\ref{fig:scharm-ctau} and~\ref{fig:braching-ratios-scharm}, respectively. From the former figure and Fig.~\ref{fig:strange-ctau},
one can see that the decay length of the ${\widetilde s_{R}}$ is about four times larger than the one of the ${\widetilde c_{R}}$, as can be easily deduced from the comparison of decay widths in Eqs.~(\ref{gamma-bnu}) and (\ref{eq:up decay}).

For the case of a ${\widetilde c_{L}}$, we obtained a lower bound for the physical mass of 1357~GeV, which corresponds to an upper bound in the decay length of 1.9~mm. Now there are two sub-dominant decay channels to quark, Higgs and neutrino/charged lepton. The contribution of each channel is $\lesssim 10^{-1}$. 
The results for the decay length and BRs 
are shown in Figs.~\ref{fig:scharm-left-ctau} and~\ref{fig:braching-ratios-scharm-left}, respectively.
Now the decay lenght is smaller than in the cases ${\widetilde s_{R}}$ and ${\widetilde c_{R}}$ because the decay width is larger due to the extra gauge contribution in Eq.~(\ref{eq:up left decay}).

Finally, for an up squark LSP in both, right and left cases, the lifetime and BRs are basically the same as for the scharm, as can be deduced from Eqs.~(\ref{eq:up decay}) and (\ref{eq:up left decay}). But the production cross section, considering the gluino mass as in Table~\ref{fixed}, is dominated by squark-squark final state, which excludes all points. In this case, it is necessary to increase the gluino mass up to about 27 TeV to decrease the contribution of this final state and allow points with up physical squark masses $\gsim$ 1800~GeV.

The above results for up-type squarks can be compared with those of a stop LSP analysed in Ref.~\cite{Kpatcha:2021nap}. In that case, 
a lower limit on the mass of the right (left) stop LSP of 1341 GeV (1068 GeV) was obtained, corresponding to a decay length of 1.86 mm (6.61 mm). \par

Let us finally remark that the use of other type of solutions for neutrino physics different from the one presented in Eq.~(\ref{neutrinomassess}), would not modify the results obtained. This can be understood from the summation over leptons present in Eqs.~\ref{gamma-bnu}-\ref{--sneutrino-decay-width-2nus2}, since for the most restrictive search \cite{CMS:2020iwv}, the results are integrated over the family. 

\section{Conclusions}
\label{sec:conclusions}
We analysed 
the signals expected at the LHC for a squark LSP of the first two families
in the framework of the $\mn$, imposing on the parameter space the experimental constraints on neutrino and Higgs physics, as well as flavour observables such as $B$ and $\mu$ decays. The squarks are pair produced and have one main two-body decay channel producing a quark and a neutrino, and two three-body sub-dominant decay modes to quark, Higgs and neutrino/charged lepton. In all cases, the three body decays represent $\lesssim 10^{-1}$ of the total decay width, and therefore their signals are not observable.

We analysed the above decay channels for two different representative values of the trilinear coupling $\lambda$ between right sneutrinos and Higgses, $\lambda\hat\nu^c\hat H_u \hat H_d$, and we found essentially the same results in both cases.
All the limits found are due to apply the LHC searches for displaced vertices of 
Ref.~\cite{CMS:2020iwv} to our model.
Because the contributions to the production cross-section of squark-squark final states are proportional to the corresponding quark density inside the proton, the results depend on the family of squarks. In particular, for $\tilde{s}_R$ LSP
we obtained a lower limit on the mass of 1646~GeV, corresponding to an upper limit on the decay length of 54.7~mm. 
For $\tilde{c}_R$ LSP, the lower limit on the mass is $1625$~GeV, whereas the upper limit on the decay length is $13.4$~mm. Finally, for $\tilde{c}_L$ LSP, the limits obtained are $1357$~GeV and $1.9$~mm.

Concerning the first family of squarks, it turns out to be excluded as LSP candidate by LHC searches, unless the gluino mass is very large. This is because to rescue points of the parameter space one has to increase the gluino mass in order to suppress the squark-squark production, so that the squark-antisquark production dominates. Thus,
to obtain a lower bound on the squark mass of about 1800 GeV, one needs
a gluino mass of about 7 TeV for down squarks and 27 TeV for up squarks.


\begin{acknowledgments}

The work of P.K. and D.L. was supported by the Argentinian CONICET, and they also acknowledge the support through
{PICT~2020-02181} and PIP11220220100320CO.  
The work of E.K. was supported by the grant "Margarita Salas" for the training of young doctors (CA1/RSUE/2021-00899), co-financed by the Ministry of Universities, the Recovery, Transformation and Resilience Plan, and the Autonomous University of Madrid.
The work of I.L.\ was funded by the Norwegian Financial Mechanism 2014-2021, grant DEC-2019/34/H/ST2/00707. 
The research of C.M. was partially supported by the AEI
through the grants IFT Centro de Excelencia Severo Ochoa No CEX2020-001007-S and PID2021-125331NB-I00, funded by MCIN/AEI/10.13039/501100011033.

\end{acknowledgments}


\appendix
\numberwithin{equation}{section}
\numberwithin{figure}{section}
\numberwithin{table}{section}

\numberwithin{equation}{subsection}
\numberwithin{figure}{subsection}
\numberwithin{table}{subsection}


\section{One Squark-two Fermion--Interactions} 
\label{appendix}

In this Appendix, we write the relevant interactions for our computation of the decays of a squark LSP of the first two families, 
following {\tt SARAH} notation~\cite{Staub:2013tta}.
In particular, 
now
$a,b=1,2,3$ are family indexes, $i,j,k$ are the indexes for the physical states, and $\alpha,\beta,\gamma=1,2,3$ are $SU(3)_C$ indexes.
The matrices $Z^D,Z^U, U^d_{L,R}, U^u_{L.R}$ and $U^V$ diagonalize the mass matrices of down squarks, up squarks, down quarks, up quarks
and neutral fermions (LH and RH neutrinos, gauginos and higgsinos), respectively.
More details about these matrices can be found in Appendix B of Ref.~\cite{Ghosh:2017yeh}.
Taking all this into account, in the basis of 4--component spinors with the projectors 
$P_{L,R}=(1\mp\gamma_5)/2$, the interactions for the mass eigenstates are as follows.

\subsection{Down squark - down quark - neutrino Interaction }

\begin{center} 
\begin{fmffile}{Figures/Diagrams/FeynDia108} 
\fmfframe(20,20)(20,20){ 
\begin{fmfgraph*}(75,75) 
\fmfleft{l1}
\fmfright{r1,r2}
\fmf{fermion}{v1,l1}
\fmf{fermion}{r1,v1}
\fmf{scalar}{r2,v1}
\fmflabel{$\bar{d}_{{i \alpha}}$}{l1}
\fmflabel{$\nu_{{j}}$}{r1}
\fmflabel{$\tilde{d}_{{k \gamma}}$}{r2}
\end{fmfgraph*}} 
\end{fmffile} 
\end{center} 

\vspace*{-0.5cm}
  \begin{align} 
\nonumber   &-\frac{i}{3} \delta_{\alpha \gamma} \Big(3 U^{V,*}_{j 6} \sum_{b=1}^{3}Z^{D,*}_{k b} \sum_{a=1}^{3}U^{d,*}_{R,{i a}} Y_{d,{a b}}    + \sqrt{2} g' U^{V,*}_{j 4} \sum_{a=1}^{3}Z^{D,*}_{k 3 + a} U^{d,*}_{R,{i a}}  \Big)P_L\\ 
    &  \,-\frac{i}{6} \delta_{\alpha \gamma} \Big(6 \sum_{b=1}^{3}\sum_{a=1}^{3}Y^*_{d,{a b}} Z^{D,*}_{k 3 + a}  U_{L,{i b}}^{d}  U_{{j 6}}^{V}  + \sqrt{2} \sum_{a=1}^{3}Z^{D,*}_{k a} U_{L,{i a}}^{d}  \Big(-3 U_{{j 5}}^{V}  + g' U_{{j 4}}^{V} \Big)\Big)P_R.
  \label{neutrinos down}
  \end{align}

\subsection
{Up squark - up quark - neutrino Interaction}


  \begin{center} 
\begin{fmffile}{Figures/Diagrams/FeynDia109} 
\fmfframe(20,20)(20,20){ 
\begin{fmfgraph*}(75,75) 
\fmfleft{l1}
\fmfright{r1,r2}
\fmf{fermion}{v1,l1}
\fmf{plain}{r1,v1}
\fmf{scalar}{r2,v1}
\fmflabel{$\bar{u}_{{i \alpha}}$}{l1}
\fmflabel{$\nu_{{j}}$}{r1}
\fmflabel{$\tilde{u}_{{k \gamma}}$}{r2}
\end{fmfgraph*}} 
\end{fmffile} 
\end{center}  

\begin{eqnarray}
  &&\frac{i}{3} \delta_{\alpha \gamma} \Big(2 \sqrt{2} g' U^{V,*}_{j 4} \sum_{a=1}^{3}Z^{U,*}_{k 3 + a} U^{u,*}_{R,{i a}}   -3 U^{V,*}_{j 7} \sum_{b=1}^{3}Z^{U,*}_{k b} \sum_{a=1}^{3}U^{u,*}_{R,{i a}} Y_{u,{a b}}   \Big)P_L \nonumber \\  
   &&-\frac{i}{6} \delta_{\alpha \gamma} \Big[6 \sum_{b=1}^{3}\sum_{a=1}^{3}Y^*_{u,{a b}} Z^{U,*}_{k 3 + a}  U_{L,{i b}}^{u}  U_{{j 7}}^{V}  + \sqrt{2} \sum_{a=1}^{3}Z^{U,*}_{k a} U_{L,{i a}}^{u}  \Big(3 g U_{{j 5}}^{V}  + g' U_{{j 4}}^{V} \Big)\Big]P_R.\nonumber\\
  \label{neutrinos up}
  \end{eqnarray}


\bibliographystyle{utphys}
\bibliography{munussmbib-completo_v6}


\end{document}